\pdfoutput=1
\documentclass[twoside,10pt]{article}
\usepackage{jmlr2e}
\usepackage{amsmath}
\usepackage{amssymb}
\usepackage{csquotes}
\hypersetup{hidelinks=true}

\usepackage{xspace}
\def\Pymanopt{\textsc{Pymanopt}\xspace}

\let\hrefold\href
\renewcommand{\href}[2]{\hrefold{#1}{\textsf{#2}}}

\usepackage{listings}
\usepackage{color}
\definecolor{deepblue}{rgb}{0,0,0.5}
\definecolor{deepred}{rgb}{0.6,0,0}
\definecolor{deepgreen}{rgb}{0,0.5,0}
\DeclareFixedFont{\ttm}{T1}{txtt}{m}{n}{10}
\newcommand\pythonstyle{\lstset{
    language=Python,
    basicstyle=\ttm,
    otherkeywords={self},
    keywordstyle=\ttm\color{deepblue},
    emph={MyClass,__init__},
    emphstyle=\ttm\color{deepred},
    stringstyle=\color{deepgreen},
    commentstyle=\ttm\color{deepgreen},
    frame=tb,
    showstringspaces=false
}}
\newcommand\pythonexternal[2][]{{
\pythonstyle
\lstinputlisting[#1]{#2}}}

\begin{document}

\title{\Pymanopt: A Python Toolbox for Optimization on Manifolds using
       Automatic Differentiation}

\author{\name James Townsend \email james.townsend.14@ucl.ac.uk\\
        \addr University College London, London, UK
        \AND
        \name Niklas Koep \email niklas.koep@rwth-aachen.de\\
        \addr RWTH Aachen University, Germany
        \AND
        \name Sebastian Weichwald \email sweichwald@tue.mpg.de\\
        \addr Max Planck Institute for Intelligent Systems, T\"ubingen, Germany
}

\editor{Antti Honkela}

\jmlrheading{17}{2016}{1-5}{4/16; Revised 7/16}{8/16}{James Townsend, Niklas Koep, Sebastian Weichwald}

\ShortHeadings{Pymanopt: A Python Toolbox for Optimization on Manifolds using Automatic Differentiation}{Townsend, Koep and Weichwald}

\maketitle

\begin{abstract}%
Optimization on manifolds is a class of methods for optimization of an
objective function, subject to constraints which are smooth, in the
sense that the set of points which satisfy the constraints admits the
structure of a differentiable manifold.
While many optimization problems are of the described form,
technicalities of differential geometry and the laborious calculation of
derivatives pose a significant barrier for experimenting with these
methods.

We introduce \Pymanopt (available at \href{https://pymanopt.github.io}{pymanopt.github.io}), a
toolbox for optimization on manifolds, implemented in Python, that---similarly
to the Manopt\footnote{Manopt is available at \href{http://manopt.org}{manopt.org}
and was introduced in \cite{manopt}.} Matlab toolbox---implements
several manifold geometries and optimization algorithms.
Moreover, we lower the barriers to users further by using automated
differentiation\footnote{We use the
term \emph{automated differentiation} to refer to the automatic calculation
of derivatives, whether using the method commonly known as \emph{automatic
differentiation}, as implemented by Autograd \citep{autograd} and TensorFlow \citep{tensorflow2015-whitepaper}, or \emph{symbolic differentiation}
as implemented by Theano \citep{team2016theano}.} for calculating derivative information,
saving users time and saving them from potential calculation and
implementation errors.
\end{abstract}

\begin{keywords}
    Riemannian optimization, non-convex optimization, manifold
    optimization, projection matrices, symmetric matrices, rotation
    matrices, positive definite matrices
\end{keywords}

\section{Introduction}

Optimization on manifolds, or Riemannian optimization, is a method for solving problems of the form
\begin{equation*}
\min_{x\in \mathcal{M}} f(x)
\end{equation*}
where $f\colon \mathcal{M}\to\mathbb{R}$ is a (cost) function and the search space
$\mathcal{M}$ is smooth, in the sense that it
admits the structure of a differentiable manifold. Although the
definition of differentiable manifold is technical and abstract, many
familiar sets satisfy this definition and are therefore compatible with
the methods of optimization on manifolds. Examples include the \emph{sphere}
(the set of points with unit Euclidean norm) in $\mathbb{R}^n$, the set of
\emph{positive definite matrices}, the set of \emph{orthogonal matrices}
as well as the set of $p$-dimensional subspaces of $\mathbb{R}^n$ with
$p < n$, also known as the \emph{Grassmann} manifold.

To perform optimization, the function $f$ needs to be defined for points on the
manifold $\mathcal{M}$. Elements of $\mathcal{M}$ are often represented by
elements of $\mathbb{R}^n$ or $\mathbb{R}^{m\times n}$, and $f$ is often well defined
on some or all of this \enquote{ambient} Euclidean space. If $f$ is also
differentiable, it makes sense for
an optimization algorithm to use the derivatives of $f$ and adapt them to the
manifold setting in order to iteratively refine solutions based on curvature
information. This is one of the key aspects of Manopt \citep{manopt}, which
allows the user to pass a function's gradient and Hessian to state of the art
solvers which exploit this information to optimize over the manifold
$\mathcal{M}$. However, working out and implementing gradients and higher
order derivatives is a laborious and error prone task, particularly when the
objective function acts on matrices or higher rank tensors. Manopt's state of
the art Riemannian Trust Regions solver, described in \cite{absil2007},
requires second order directional derivatives (or a
numerical approximation thereof), which are particularly challenging to work out
for the average user, and more error prone and tedious even for an experienced
mathematician.

It is these difficulties which we seek to address with this toolbox.
\Pymanopt supports a variety of modern Python libraries for
automated differentiation of cost functions acting on vectors, matrices or
higher rank tensors. Combining optimization on manifolds and
automated differentiation enables a convenient workflow for rapid
prototyping that was previously unavailable to practitioners. All that is
required of the user is to instantiate a manifold, define a cost function, and
choose one of \Pymanopt's solvers. This means that the Riemannian Trust
Regions solver in \Pymanopt is just as easy to use as one of the
derivative-free or first order methods.

\section{The Potential of Optimization on Manifolds and \Pymanopt Use Cases}

Much of the theory of how to adapt Euclidean optimization algorithms to
(matrix) manifolds can be found in \cite{smith1994, edelman1998, absil2008}.
The approach of optimization on manifolds is superior to performing free (Euclidean)
optimization and projecting the parameters back onto the
search space after each iteration (as in the projected gradient
descent method), and has been shown to outperform standard algorithms
for a number of problems.

\cite{hosseini2015} demonstrate this advantage for a well-known problem in
machine learning, namely inferring the maximum likelihood parameters of a
mixture of Gaussian (MoG) model.  Their alternative to the
traditional expectation maximization (EM) algorithm uses
optimization over a product manifold of positive definite (covariance)
matrices. Rather than optimizing the likelihood function directly, they
optimize a reparameterized version which shares the same local optima.  The
proposed method, which is on par with EM and shows less variability in running
times, is a striking example why we think a toolbox like \Pymanopt, which
allows the user to readily experiment with and solve problems involving
optimization on manifolds, can
accelerate and pave the way for improved machine learning algorithms.\footnote{A quick example implementation for inferring MoG parameters
is available at
\href{https://pymanopt.github.io/MoG.html}{pymanopt.github.io/MoG.html}.}

Further successful applications of optimization on manifolds include matrix
completion tasks \citep{vandereycken2013, boumal2015},
robust PCA \citep{podosinnikova2014},
dimension reduction for
independent component analysis (ICA) \citep{theis2009}, kernel ICA
\citep{shen2007} and similarity learning \citep{shalit2012}.

Many more applications to machine learning and other fields exist. While a full
survey on the usefulness of these methods is well beyond the scope
of this manuscript, we highlight that at the time of writing, a search for the
term \enquote{manifold optimization} on the IEEE Xplore Digital Library lists
1065 results; the Manopt toolbox itself is referenced in 90 papers indexed by
Google Scholar.

\section{Implementation}
Our toolbox is written in Python and uses NumPy and SciPy for
computation and linear algebra operations. Currently \Pymanopt is
compatible with cost functions defined using
Autograd \citep{autograd},
Theano \citep{team2016theano}
or
TensorFlow \citep{tensorflow2015-whitepaper}.
\Pymanopt itself and all the required software is open source, with no
dependence on proprietary software.

To calculate derivatives, Theano uses symbolic differentiation, combined with
rule-based optimizations,
while both Autograd and TensorFlow use reverse-mode
automatic differentiation. For a
discussion of the distinctions between the two approaches and an overview of
automatic differentiation in the context of machine learning, we refer the
reader to \cite{baydin2015}.

Much of the structure of \Pymanopt is based on that of the Manopt Matlab
toolbox. For this early release, we have implemented  all of the solvers and a
number of the manifolds found in Manopt, and plan to implement more, based on
the needs of users. The codebase is structured in a modular way and thoroughly
commented to make extension to further solvers, manifolds, or backends for automated differentiation
as straightforward as possible. Both a user and developer documentation are
available. The GitHub
repository at \href{https://github.com/pymanopt/pymanopt}{github.com/pymanopt/pymanopt}
offers a convenient way to ask for help or request features by raising an
issue, and contains guidelines for those wishing to contribute to the project.

\section{Usage: A Simple Instructive Example}

All automated differentiation in \Pymanopt is performed behind the scenes so
that the amount of setup code required by the user is minimal. Usually only the
following steps are required:
\begin{enumerate}\itemsep0em
    \item[(a)] Instantiation of a manifold $\mathcal{M}$
    \item[(b)] Definition of a cost function $f\colon\mathcal{M}\to\mathbb{R}$
    \item[(c)] Instantiation of a \Pymanopt solver
\end{enumerate}

We briefly demonstrate the ease of use with a simple example.
Consider the problem of finding an $n\times n$ positive semi-definite (PSD)
matrix $S$ of rank $k<n$ that best approximates a given $n\times n$
(symmetric) matrix $A$, where closeness between $A$ and its low-rank
PSD approximation $S$ is measured by the following loss function
$$L_{\delta}(S,A) \triangleq \sum_{i=1}^n \sum_{j=1}^n H_\delta\left(s_{i,j}-a_{i,j}\right)$$
for some $\delta > 0$ and $H_\delta(x) \triangleq \sqrt{x^2+\delta^2}-\delta $ the pseudo-Huber loss function.
This loss function is robust against outliers as $H_\delta(x)$ approximates $|x|-\delta$ for large values of $x$ while being approximately quadratic for small values of $x$ \citep{huber1964robust}.

This can be formulated as an optimization problem on the manifold of PSD matrices:
$$\min_{S\in\mathcal{PSD}_k^n} L_{\delta}(S,A)$$
where $\mathcal{PSD}_k^n\triangleq\{M\in\mathbb{R}^{n\times n}:M\succeq 0,
\operatorname{rank}(M)=k\}$. This task is easily solved using \Pymanopt:

\pythonexternal{example.py}

The examples folder within the \Pymanopt toolbox holds further instructive examples, such as performing inference in mixture of Gaussian models using optimization on manifolds instead of the expectation maximization algorithm. Also see the examples section on \href{http://pymanopt.github.io}{pymanopt.github.io}.

\section{Conclusion}
\Pymanopt enables the user to experiment with different state
of the art solvers for optimization problems on manifolds, like the Riemannian Trust Regions
solver, without any extra effort. Experimenting with different cost functions,
for example by changing the pseudo-Huber loss $L_{\delta}(S,A)$ in the code above to the Frobenius norm $||S-A||_F$, a $p$-norm $||S-A||_p$, or some more complex function, requires just a small change in the definition of the cost
function. For problems of greater complexity, \Pymanopt offers a significant
advantage over toolboxes that require manual differentiation by enabling users
to run a series of related experiments without returning
to pen and paper each time to work out derivatives. Gradients and Hessians only
need to be derived if they are required for other analysis of a problem. We
believe that these advantages, coupled with the potential for extending
\Pymanopt to large-scale applications using TensorFlow, could lead to
significant progress in applications of optimization on manifolds.

\section*{Acknowledgments}
We would like to thank the developers of the Manopt Matlab toolbox, in
particular Nicolas Boumal and Pierre-Antoine Absil, for developing Manopt,
and for the generous help and advice they have given. We would also like
to thank Heiko Strathmann for his thoughtful advice as well as the anonymous reviewers for their constructive feedback and idea for a more suitable application example.

\newpage

\bibliography{pymanopt_paper}

\begin{thebibliography}{17}
\providecommand{\natexlab}[1]{#1}
\providecommand{\url}[1]{\texttt{#1}}
\expandafter\ifx\csname urlstyle\endcsname\relax
  \providecommand{\doi}[1]{doi: #1}\else
  \providecommand{\doi}{doi: \begingroup \urlstyle{rm}\Url}\fi

\bibitem[Abadi et~al.(2015)Abadi, Agarwal, Barham, Brevdo, Chen, Citro,
  Corrado, Davis, Dean, Devin, Ghemawat, Goodfellow, Harp, Irving, Isard, Jia,
  Jozefowicz, Kaiser, Kudlur, Levenberg, Man\'{e}, Monga, Moore, Murray, Olah,
  Schuster, Shlens, Steiner, Sutskever, Talwar, Tucker, Vanhoucke, Vasudevan,
  Vi\'{e}gas, Vinyals, Warden, Wattenberg, Wicke, Yu, and
  Zheng]{tensorflow2015-whitepaper}
M.~Abadi, A.~Agarwal, P.~Barham, E.~Brevdo, Z.~Chen, C.~Citro, G.~S. Corrado,
  A.~Davis, J.~Dean, M.~Devin, S.~Ghemawat, I.~Goodfellow, A.~Harp, G.~Irving,
  M~Isard, Y.~Jia, R.~Jozefowicz, L.~Kaiser, M.~Kudlur, J.~Levenberg,
  D.~Man\'{e}, R.~Monga, S.~Moore, D.~Murray, C.~Olah, M.~Schuster, J.~Shlens,
  B.~Steiner, I.~Sutskever, K.~Talwar, P.~Tucker, V.~Vanhoucke, V.~Vasudevan,
  F.~Vi\'{e}gas, O.~Vinyals, P.~Warden, M.~Wattenberg, M.~Wicke, Y.~Yu, and
  X.~Zheng.
\newblock {TensorFlow: Large-Scale Machine Learning on Heterogeneous Systems},
  2015.
\newblock URL \url{http://tensorflow.org}.

\bibitem[Absil et~al.(2007)Absil, Baker, and Gallivan]{absil2007}
P.-A. Absil, C.G. Baker, and K.A. Gallivan.
\newblock Trust-{R}egion {M}ethods on {R}iemannian {M}anifolds.
\newblock \emph{Foundations of Computational Mathematics}, 7\penalty0
  (3):\penalty0 303--330, 2007.

\bibitem[Absil et~al.(2008)Absil, Mahony, and Sepulchre]{absil2008}
P.-A. Absil, R.~Mahony, and R.~Sepulchre.
\newblock \emph{Optimization Algorithms on Matrix Manifolds}.
\newblock Princeton University Press, Princeton, NJ, 2008.
\newblock ISBN 978-0-691-13298-3.

\bibitem[Al-Rfou et~al.(2016)Al-Rfou, Alain, Almahairi, Angermueller, Bahdanau,
  Ballas, Bastien, Bayer, Belikov, Belopolsky, Bengio, Bergeron, Bergstra,
  Bisson, {Bleecher Snyder}, Bouchard, Boulanger-Lewandowski, Bouthillier,
  de~Br\'ebisson, Breuleux, Carrier, Cho, Chorowski, Christiano, Cooijmans,
  C\^ot\'e, C\^ot\'e, Courville, Dauphin, Delalleau, Demouth, Desjardins,
  Dieleman, Dinh, Ducoffe, Dumoulin, {Ebrahimi Kahou}, Erhan, Fan, Firat,
  Germain, Glorot, Goodfellow, Graham, Gulcehre, Hamel, Harlouchet, Heng,
  Hidasi, Honari, Jain, Jean, Jia, Korobov, Kulkarni, Lamb, Lamblin, Larsen,
  Laurent, Lee, Lefrancois, Lemieux, L\'eonard, Lin, Livezey, Lorenz, Lowin,
  Ma, Manzagol, Mastropietro, McGibbon, Memisevic, van Merri\"enboer,
  Michalski, Mirza, Orlandi, Pal, Pascanu, Pezeshki, Raffel, Renshaw, Rocklin,
  Romero, Roth, Sadowski, Salvatier, Savard, Schl\"uter, Schulman, Schwartz,
  Serban, Serdyuk, Shabanian, Simon, Spieckermann, Subramanyam, Sygnowski,
  Tanguay, van Tulder, Turian, Urban, Vincent, Visin, de~Vries, Warde-Farley,
  Webb, Willson, Xu, Xue, Yao, Zhang, and Zhang]{team2016theano}
R.~Al-Rfou, G.~Alain, A.~Almahairi, C.~Angermueller, D.~Bahdanau, N.~Ballas,
  F.~Bastien, J.~Bayer, A.~Belikov, A.~Belopolsky, Y.~Bengio, A.~Bergeron,
  J.~Bergstra, V.~Bisson, J.~{Bleecher Snyder}, N.~Bouchard,
  N.~Boulanger-Lewandowski, X.~Bouthillier, A.~de~Br\'ebisson, O.~Breuleux,
  P.-L. Carrier, K.~Cho, J.~Chorowski, P.~Christiano, T.~Cooijmans, M.-A.
  C\^ot\'e, M.~C\^ot\'e, A.~Courville, Y.N. Dauphin, O.~Delalleau, J.~Demouth,
  G.~Desjardins, S.~Dieleman, L.~Dinh, M.~Ducoffe, V.~Dumoulin, S.~{Ebrahimi
  Kahou}, D.~Erhan, Z.~Fan, O.~Firat, M.~Germain, X.~Glorot, I.~Goodfellow,
  M.~Graham, C.~Gulcehre, P.~Hamel, I.~Harlouchet, J.-P. Heng, B.~Hidasi,
  S.~Honari, A.~Jain, S.~Jean, K.~Jia, M.~Korobov, V.~Kulkarni, A.~Lamb,
  P.~Lamblin, E.~Larsen, C.~Laurent, S.~Lee, S.~Lefrancois, S.~Lemieux,
  N.~L\'eonard, Z.~Lin, J.~A. Livezey, C.~Lorenz, J.~Lowin, Q.~Ma, P.-A.
  Manzagol, O.~Mastropietro, R.T. McGibbon, R.~Memisevic, B.~van Merri\"enboer,
  V.~Michalski, M.~Mirza, A.~Orlandi, C.~Pal, R.~Pascanu, M.~Pezeshki,
  C.~Raffel, D.~Renshaw, M.~Rocklin, A.~Romero, M.~Roth, P.~Sadowski,
  J.~Salvatier, F.~Savard, J.~Schl\"uter, J.~Schulman, G.~Schwartz, I.V.
  Serban, D.~Serdyuk, S.~Shabanian, \'E. Simon, S.~Spieckermann, S.R.
  Subramanyam, J.~Sygnowski, J.~Tanguay, G.~van Tulder, J.~Turian, S.~Urban,
  P.~Vincent, F.~Visin, H.~de~Vries, D.~Warde-Farley, D.J. Webb, M.~Willson,
  K.~Xu, L.~Xue, L.~Yao, S.~Zhang, and Y.~Zhang.
\newblock {Theano: A {Python} framework for fast computation of mathematical
  expressions}.
\newblock \emph{arXiv preprint arXiv:1605.02688}, 2016.
\newblock URL \url{http://deeplearning.net/software/theano}.

\bibitem[Baydin et~al.(2015)Baydin, Pearlmutter, Radul, and
  Siskind]{baydin2015}
A.G. Baydin, B.A. Pearlmutter, A.A. Radul, and J.M. Siskind.
\newblock Automatic differentiation in machine learning: a survey.
\newblock \emph{arXiv preprint arXiv:1502.05767}, 2015.

\bibitem[Boumal and Absil(2015)]{boumal2015}
N.~Boumal and P.-A. Absil.
\newblock Low-rank matrix completion via preconditioned optimization on the
  {G}rassmann manifold.
\newblock \emph{Linear Algebra and its Applications}, 475:\penalty0 200--239,
  2015.
\newblock \doi{10.1016/j.laa.2015.02.027}.

\bibitem[Boumal et~al.(2014)Boumal, Mishra, Absil, and Sepulchre]{manopt}
N.~Boumal, B.~Mishra, P.-A. Absil, and R.~Sepulchre.
\newblock {{M}anopt, a {M}atlab Toolbox for Optimization on Manifolds}.
\newblock \emph{Journal of Machine Learning Research}, 15:\penalty0 1455--1459,
  2014.
\newblock URL \url{http://manopt.org}.

\bibitem[Edelman et~al.(1998)Edelman, Arias, and Smith]{edelman1998}
A.~Edelman, T.A. Arias, and S.T. Smith.
\newblock {The Geometry of Algorithms with Orthogonality Constraints}.
\newblock \emph{SIAM J. Matrix Anal. \& Appl.}, 20\penalty0 (2):\penalty0
  303--353, 1998.

\bibitem[Hosseini and Sra(2015)]{hosseini2015}
R.~Hosseini and S.~Sra.
\newblock {Matrix Manifold Optimization for Gaussian Mixtures}.
\newblock In \emph{Advances in Neural Information Processing Systems}, pages
  910--918, 2015.

\bibitem[Huber(1964)]{huber1964robust}
P.J. Huber.
\newblock Robust estimation of a location parameter.
\newblock \emph{The Annals of Mathematical Statistics}, 35\penalty0
  (1):\penalty0 73--101, 1964.

\bibitem[Maclaurin et~al.(2015)Maclaurin, Duvenaud, Johnson, and
  Adams]{autograd}
D.~Maclaurin, D.~Duvenaud, M.~Johnson, and R.P. Adams.
\newblock {Autograd: Reverse-mode differentiation of native {P}ython}, 2015.
\newblock URL \url{http://github.com/HIPS/autograd}.

\bibitem[Podosinnikova et~al.(2014)Podosinnikova, Setzer, and
  Hein]{podosinnikova2014}
A.~Podosinnikova, S.~Setzer, and M.~Hein.
\newblock {Robust PCA: Optimization of the Robust Reconstruction Error over the
  Stiefel Manifold}.
\newblock In \emph{36th German Conference on Pattern Recognition (GCPR)}, 2014.

\bibitem[Shalit et~al.(2012)Shalit, Weinshall, and Chechik]{shalit2012}
U.~Shalit, D.~Weinshall, and G.~Chechik.
\newblock {Online Learning in the Embedded Manifold of Low-rank Matrices}.
\newblock \emph{Journal of Machine Learning Research}, 13\penalty0
  (1):\penalty0 429--458, 2012.

\bibitem[Shen et~al.(2007)Shen, Jegelka, and Gretton]{shen2007}
H.~Shen, S.~Jegelka, and A.~Gretton.
\newblock {Fast Kernel ICA using an Approximate Newton Method}.
\newblock In \emph{International Conference on Artificial Intelligence and
  Statistics}, pages 476--483, 2007.

\bibitem[Smith(1994)]{smith1994}
S.T. Smith.
\newblock {Optimization techniques on Riemannian manifolds}.
\newblock \emph{Fields institute communications}, 3\penalty0 (3):\penalty0
  113--135, 1994.

\bibitem[Theis et~al.(2009)Theis, Cason, and Absil]{theis2009}
F.J. Theis, T.P. Cason, and P.-A. Absil.
\newblock {Soft dimension reduction for ICA by joint diagonalization on the
  Stiefel manifold}.
\newblock In \emph{Independent Component Analysis and Signal Separation}, pages
  354--361. Springer, 2009.

\bibitem[Vandereycken(2013)]{vandereycken2013}
B.~Vandereycken.
\newblock {Low-Rank Matrix Completion by Riemannian Optimization}.
\newblock \emph{SIAM J. Optim.}, 23\penalty0 (2):\penalty0 1214--1236, 2013.

\end{thebibliography}

\end{document}